\documentclass[aps,preprint,superscriptaddress,]{revtex4}
\usepackage{amsfonts}
\usepackage{amsmath}
\usepackage{amssymb}
\usepackage{graphicx}
\usepackage{mathrsfs}

\begin{document}
\title{Comment on ``What the information loss is {\it not}"}
\author{Qing-yu Cai}
\email{qycai@wipm.ac.cn} \affiliation{State Key Laboratory of
Magnetic Resonances and Atomic and Molecular Physics, Wuhan
Institute of Physics and Mathematics, Chinese Academy of Sciences,
Wuhan 430071, People's Republic of China}
\author{Baocheng Zhang}
%\email{zhangbc@wipm.ac.cn}
\affiliation{State Key Laboratory of
Magnetic Resonances and Atomic and Molecular Physics,
Wuhan Institute of Physics and Mathematics, Chinese Academy of Sciences,
Wuhan 430071, People's Republic of China}
\author{Ming-Sheng Zhan}
%\email{mszhan@wipm.ac.cn}
\affiliation{State Key Laboratory of
Magnetic Resonances and Atomic and Molecular Physics, Wuhan
Institute of Physics and Mathematics, Chinese Academy of Sciences,
Wuhan 430071, People's Republic of China}
\author{Li You}
\email{lyou@mail.tsinghua.edu.cn}
\affiliation{State Key Laboratory
of Low Dimensional Quantum Physics, Department of Physics, Tsinghua
University, Beijing 100084, People's Republic of China}


\begin{abstract}

A recent article by Mathur attempts a ``precise formulation" for the paradox
of black hole information loss [S. D. Mathur, arXiv:1108.0302v2 (hep-th)].
We point out that a key component of the above work,
which refers to entangled pairs inside and outside of the horizon
and their associated entropy gain or information loss during black hole
evaporation, is a presumptuous false outcome not backed by
the very foundation of physics.
The very foundation of Mathur's above work is thus incorrect.
We further show that within the frame work of
Hawking radiation as tunneling the so-called small corrections
are sufficient to resolve the information loss problem.

\end{abstract}

\volumeyear{year}
\volumenumber{number}
\issuenumber{number}
\eid{identifier}
\date{\today}
\maketitle

\section{The black hole information loss paradox}

The black hole information loss paradox can be appreciated without
getting into the details of quantum gravity.
As first addressed by Hawking \cite{haw}, the information loss paradox,
or the Hawking paradox, refers to the ever increasing entropy
of the thermal radiations (Hawking radiations) from a black hole.
If the radiations are thermal (thermally distributed) with no correlations
between individual radiated quanta, an increasing entropy equals to
loss of information.  In order to resolve the paradox,
one must find a mechanism capable of balance the entropy increase.
Many scenarios for possible resolutions have been discussed at varying
levels of details. For instance, we recently point out the increasing
entropy is balanced exactly by the nontrivial correlations among
Hawking radiations if the non-thermal spectrum based on
Hawking radiation as tunneling is adopted.

The information-carrying correlation we discover will play a crucial
role in any resolution. Yet, the Hawking paradox cannot be
considered resolved even if these correlations are detected. A
complete resolution can only occur after entropy conservation is
confirmed, which then implies the dynamics of Hawking radiation is
consistent with unitarity as required by quantum mechanics, despite
our inabilities in writing down a time-dependent wave function for
the emissions consistent with quantum statistics and general
relativity.

\section{Mathur's ``precise formulation'' of the paradox}

Recently, Mathur concludes that small corrections cannot resolve the
information loss paradox \cite{sm10}. His reasoning rests upon the so-called
``precise formulation" of the paradox, outlined in four steps A-D \cite{sm11}.
In his step (C) ``Creation of the entangled state",
EPR pairs of the type
\begin{equation}
|\Psi\rangle_{\rm pair}=\frac{1}{\sqrt{2}}
(|0\rangle_{c_i}|0\rangle_{b_i}+|1\rangle_{c_i}|1\rangle_{b_i}),
\label{epr}
\end{equation}
are assumed to be created one after another near the horizon
with evolution. After $N$ pairs the whole
wave function then takes the following form
\begin{equation}
|\Psi\rangle\approx |\Psi\rangle_M\otimes
(\frac{1}{\sqrt{2}}|0\rangle_{c_i}|0\rangle_{b_i}
+\frac{1}{\sqrt{2}}|1\rangle_{c_i}|1\rangle_{b_i})^{\otimes N},
\label{state}
\end{equation}
where $|\Psi\rangle_M $ is the state of the matter shell.
The regions outside the horizon are labelled $b_i$ while
those inside the horizon are labelled $c_i$.
the numbers $0$ and $1$ denote the occupation number of a particular mode.
The mass of a black hole decreases by the ingoing (negative massed) particle $c_i$
with each (positive massed) particle $b_i$ outside the horizon
becoming Hawking radiation. In the end, the hole vanishes due to exhaustion.

Mathur concludes the initial pure state $|\Psi\rangle_M$ for the
infalling hole evolves into a mixture claiming it is not entangled with
emissions $b_i$, outside the horizon. $b_i$ is therefore
left also in a mixed state as there is nothing remaining in the hole
for it to be entangled with. This is clearly erroneous as
we show in detail below.
When the initial correlation between $c_i$ and $b_i$
is considered as Mathur does in his `precise formulation',
$b_i$ actually evolves into a pure state upon
the annihilation of $c_i$ into the black hole.
Rather than relying on unstated assumptions or philosophical arguments,
we will walk through Mathur's scenario line by line, step by step,
exam his mistake in detail.

Let us consider the EPR pair state
$|\Psi\rangle_{\rm pair}=(|0\rangle_{c_i}|0\rangle_{b_i}
+|1\rangle_{c_i}|1\rangle_{b_i})/\sqrt{2}$ for the particles
$b_i$ and $c_i$ respectively outside and inside the horizon.
The state for particle $b_{i}$ is found to be $\rho_{b_i}=(|0\rangle\langle
0|+|1\rangle\langle 1|)/2$ after tracing out $c_i$, which is a mixed
state, actually a maximally mixed state. Likewise, we find $\rho_{c_i}=(|0\rangle\langle
0|+|1\rangle\langle 1|)/2$. The respective entropies for the
$b_i$ and $c_i$ are
\begin{equation}
S(b_i)= S(c_i)=\ln2,
\label{entropy}
\end{equation}
which are nonzero as both are in mixed states.
However, the EPR pair state $|\Psi\rangle_{\rm pair}$ is a pure state,
whose entropy is simply zero, or $S(|\Psi\rangle\langle \Psi|_{\rm pair})=0$.
This helps us to obtain the entropy of the correlation between $b_i$ and $c_i$
\begin{equation}
S(b_i:c_i)=-2\ln2,
\label{correlation}
\end{equation}
as $S(b_i)+S(c_i)+S(b_i:c_i)=0$.

When a negative energy (mass $m_{c_i}=-\omega$) particle $c_i$ falls
into the hole, it eventually annihilates a positive energy matter
$a_i$ of $m_{a_i}=\omega$. The mass of the hole thus decreases by
$-\omega$ since $m_{a_i}+m_{c_i}=0$. The annihilation actually
causes the particles $a_{i}$ and $c_{i}$ to end up in a maximally
entangled state, a point presumptuously skipped over in Mathur's
reformulation. This is the reason for his subsequent false statement
and conclusion. In essence, the annihilation is like a joint
measurement in teleportation, causing the state of particle $a_{i}$
to be teleported to the emission $b_{i}$ outside the horizon
\cite{b93}. If $a_i$ initially exists in a pure state, $b_i$ will
end up in a pure state, not a mixed state as Mathur concluded,
although $b_i$ is no longer entangled with $c_i$ or $a_i$ after
their annihilation inside the horizon.

To be more precise, the joint state $|\Psi\rangle$ for the whole
black hole including EPR pairs cannot be written down in a product
state Eq. (\ref{state}) because its temperature changes after each
emission $\omega$, {\it i.e.}, $M\rightarrow M-\omega_i$. Taking
this into account, the joint state is more appropriately expressed
as
\begin{eqnarray}
|\Psi\rangle=\lambda_{i}|a_i\rangle|c_i\rangle|b_i\rangle,
\label{joint}
\end{eqnarray}
where $\lambda_{i}\sim e^{-4\pi M_{i}\omega_{i}}$ denotes the
temperature change due to an emission $b_i$.
When the hole is exhausted due to annihilations
of $a_i$ and $c_i$ pairs, the initial pure state of the hole,
$|\Psi\rangle_{M}$,  is teleported to the Harking radiations.
A detailed discussion of the above is already present and can be found
in Ref. \cite{sl06}, where one finds the final state of Hawking radiation
is a pure state [see Eq. (9) of \cite{sl06}], not a mixed state as
assumed by Mathur.

We thus claim Mathur's conclusion that the initial pure state
$|\Psi\rangle_{M}$ will involve into a mixed state because
the quanta $b_i$ outside the horizon is in a mixed state is false.

\section{small corrections are sufficient}

The temperature of a black hole is inversely proportional to its mass,
thus it is most easily described in terms of an ensemble with
its entropy, $S_{\mathrm{BH}}=4\pi M^2$,
or the Bekenstein entropy for a Schwarzschild black hole \cite{Ben}.
Although it is possible in principle to carry out a pure state analysis
of the Hawking radiation dynamics, we do not know how to write down
an appropriate initial pure state. The statistical properties of a black hole,
on the other hand,
can be described once its entropy is known.
After the hole is vanished into Hawking radiations,
the entropy for the remaining system made up of emissions becomes
$S_{\rm emissions}=\sum_{i}S(\omega_{i})$
with $M=\sum_{i}\omega_{i}$ due to energy conservation.
The information loss paradox arises as a result of total entropy increase,
or $S_{\rm emissions}>S_{\mathrm{BH}}$,
if the spectrum for Hawking radiation is thermal.

Based on an earlier work by us \cite{zc09,zc11}, it is straightforward
working backwards to find out what kind of spectrum for Hawking radiation
is needed in order to resolve the paradox making use of the information
carrying correlations between emissions.
This rests in our belief of unitarity in quantum mechanics,
{\it i.e.}, information is conserved,
which requires the total entropy of a closed system,
composed of the hole and the emissions in the present case, to be conserved.
After an emission of energy $\omega$, the entropy for the black hole
decreases by $\Delta S_{\mathrm{BH}}=4\pi (M-\omega)^2-4\pi M^2=-8\pi M\omega+4\pi
\omega^2$. From $\Delta S_{\mathrm{BH}}=\ln\Gamma(\omega)$, we find the
appropriate spectrum is $\Gamma(\omega)=\exp[-8\pi\omega(M-\omega/2)]$,
or precisely the nonthermal spectrum of Parikh and Wilczek \cite{pw00}.
When compared to Hawking's thermal spectrum, it contains a small correction
$O(\omega^2)$, due to the back action of the quantum emission event. Although
tiny in magnitude for most massive black holes, this small correction
in the Parikh-Wilczek spectrum \cite{pw00} provides exactly what is needed to
balance the total entropy increase with information contained in
correlations among Hawking radiations.
Thus, due to our discovery of the non-trivial correlations
among radiations governed by the Parikh-Wilczek spectrum \cite{zc09},
one can now conclude that small corrections
in the nonthermal spectrum are sufficient to resolve the paradox
of black hole information loss.

\section{Mathur's criticism on our resolution}

As outlined above, the spectrum of Parikh and Wilczek,
based on Hawking radiation as tunneling, is sufficient
to resolve the information loss paradox.
In the article by Mathur \cite{sm11}, however,
a criticism is raised regarding our resolution (see page 14 of Ref. \cite{sm11}).

Mathur argues that there exists a missing prefactor in the Parikh-Wilczek
spectrum $\Gamma(\omega)=\exp[-8\pi\omega(M-\omega/2)]$, which should have
the appropriate dimension of frequency like an attempt frequency in
the standard tunneling treatment for a particle of mass $\omega$.
Thus $\Gamma(\omega)$ represents the probability flux of transmitted
particles. This point of a missing prefactor is certainly correct.
Generally speaking, $\Gamma=\Lambda\exp[-8\pi\omega(M-\omega/2)]$
indeed comes with a prefactor $\Lambda$.
When the probability flux of the incident particles is assumed unity,
the prefactor becomes $\Lambda=1$.

The key criticism of Mathur to our resolution concerns
the crucial relationship used by us for the nonthermal spectrum,
\begin{equation}
\Gamma(\omega_1,\omega_2)=\Gamma(\omega_1)\Gamma(\omega_2|\omega_1)=\Gamma(\omega_1+\omega_2).
\label{relation}
\end{equation}
Mathur proclaims this to be true even for the thermal spectrum
$\Gamma_{T}(\omega)=\exp(-8\pi M\omega)$, which
is a totaly erroneous as we show below.
Using statistical theory,
it is easy to find for the thermal spectrum the
various terms are given by
$\Gamma(\omega_2|\omega_1)=\exp[-8\pi (M-\omega_1)\omega_2]$,
$\Gamma(\omega_1,\omega_2)=\Gamma(\omega_1)\Gamma(\omega_2|\omega_1)=\exp[-8\pi
M(\omega_1+\omega_2)]\exp(8\pi\omega_1\omega_2)$, and
$\Gamma(\omega_1+\omega_2)=\exp[-8\pi M(\omega_1+\omega_2)]$.
Thus it is easy to check that
$\Gamma(\omega_1,\omega_2)=\Gamma(\omega_1+\omega_2)\exp(8\pi\omega_1\omega_2)\neq
\Gamma(\omega_1+\omega_2)$ .
Therefore, this key criticism from Mathur to our resolution
is based on a rather elementary mistake in his calculation of the conditional
probability and joint probability. His criticism cannot stand.

Finally, we stress that the essential difference between the
thermal spectrum of Hawking and the nonthermal spectrum of Parikh-Wilczek
translates into the absence and existence of correlations among different emissions.
Although rather small between any two individual emissions, the correlations
among all pairs of Hawking radiations have the capacity to carry off
all information in the initial black hole, thus the information loss
paradox is resolved \cite{iy10}.

\section{conclusion}

In summary, we point out that Mathur's ``precise formulation"
of the information loss paradox lacks content and is false in
several of the key components.
His key calculations are incorrect and his conclusion
regarding our resolution based on Hawking radiation
as tunneling is erroneous.
In the simply example he provides to discuss entanglement and
correlation between two particles in an EPR pair respectively
residing inside and outside the horizon, the final state
of emissions is a pure state when quantum correlations
are properly treated, again in stark contrast to his presumptuous
statement of a mixture state.

We further point out that
Mathur's criticism on our resolution of Hawking radiation
as tunneling is once again due to an incorrect use of
probability theory.
We show using straightforward statistics and probability theory that
there exists information-carrying correlations among
Hawking radiations \cite{zc09,zc11,iy10} and the entropy is
conserved during Hawking radiation process, when the non-thermal
spectrum of Parikh and Wilczek is assumed, or when back reaction
from Hawking radiation is considered \cite{pw00}.
Our discovery thus proves undoubtedly that no information is lost in
Hawking radiation. The small correction in the spectrum of
Parikh and Wilczek is necessary and sufficient to resolve the
information loss paradox \cite{zc09,zc11,iy10}.

\end{document}